\documentclass{article}
\usepackage{waspaa23,amsmath,graphicx,url,times}
\usepackage{color}
\usepackage{enumitem}
\usepackage{hyperref}
\usepackage{url}
\usepackage{breakurl}
\usepackage{multirow,makecell}
\usepackage{etoolbox,siunitx}
\robustify\bfseries
\usepackage{booktabs}
\usepackage{balance}
\usepackage{mathrsfs}
\usepackage{amsfonts}
\usepackage{bm}
\usepackage{arydshln}
\usepackage{caption}
\usepackage{nicefrac}
\usepackage{cite}
\usepackage{cite}
\usepackage{bbm}
\usepackage[scr=boondoxo]{mathalfa}

\title{Complete and separate: Conditional separation\\with missing target source attribute completion}

\name{Dimitrios Bralios, Efthymios Tzinis, Paris Smaragdis}
\address{University of Illinois Urbana-Champaign, Department of Computer Science, Urbana, IL, USA}
\begin{document}

\ninept
\maketitle

\begin{sloppy}

\begin{abstract}


Recent approaches in source separation leverage semantic information about their input mixtures and constituent sources that when used in conditional separation models can achieve impressive performance. Most approaches along these lines have focused on simple descriptions, which are not always useful for varying types of input mixtures. In this work, we present an approach in which a model, given an input mixture and partial semantic information about a target source, is trained to extract additional semantic data. We then leverage this pre-trained model to improve the separation performance of an uncoupled multi-conditional separation network. Our experiments demonstrate that the separation performance of this multi-conditional model is significantly improved, approaching the performance of an oracle model with complete semantic information. Furthermore, our approach achieves performance levels that are comparable to those of the best performing specialized single conditional models, thus providing an easier to use alternative.

\end{abstract}


\section{Introduction}
\label{sec:intro}
Undoubtedly, sound source separation is one of the most fundamental problems in audio scene analysis \cite{wang2006computational}. Neural network training recipes like deep clustering and permutation invariant training (PIT) \cite{hershey2016deepclustering,Isik2016Interspeech09,Yu2017PIT} have achieved remarkable performance in speech, music and universal sound separation tasks by separating all independent source waveforms from an input mixture recording \cite{ZQwang2022tf}. However, slicing an acoustic scene according to the user's intent might also require models which are aware of other discriminative attributes of the mixture's constituent sources in order to specify a target waveform using conditional information \cite{schulze2019weakly}. Earlier works have analyzed the importance of conditionally-informed separation models by estimating class-labels of the mixture's constituent sources \cite{tzinis2020improving,chen2014feature} and using event-based queries for speech enhancement \cite{xin2023improving}. Further, some have explored two-stage approaches, where the stage preceding separation extracts semantic information about the sources \cite{zeghidour2021wavesplit}. Conversely, others have shown that sound event detection systems can benefit from sound separation front-ends \cite{turpault2020improvingSoundEventDetectionusingSeparation}. 

One straightforward way to explicitly condition separation models \cite{meseguer2019conditioned} in order to target a source of interest is to use features related to a speaker's identity \cite{wang2019voicefilter} or to describe sounds with class-based semantics \cite{ochiai20_interspeech}. Other types of condition vectors or queries might also contain text-based descriptions \cite{liu22w_interspeech,kilgour22_interspeech,dong2023clipsep} or visual-cues for on-screen audio-visual separation \cite{tzinis2022audioscopev2}. Although the aforementioned methods have successfully proposed single-condition models, more recent works have also proposed to train conditional separation models by combining different and probably non-mutually exclusive discriminative concepts \cite{tzinis22_interspeech,tzinis2022optimal}. The proposed heterogeneous condition training (HCT) \cite{tzinis22_interspeech} recipe is able to produce a more general multi-condition separation model while sometimes surpassing unconditional separation PIT models by leveraging the complimentarity of information originating from the different condition vectors to describe one target source. On the other hand, the performance of HCT is inferior compared to specialist single-conditioned models when the number and/or the difficulty of the conditions that need to be learned increases. To that end, optimal condition training (OCT) \cite{tzinis2022optimal} has been proposed in order to bridge this separation performance gap by performing gradient steps towards the maximal performing condition vector. As shown in \cite{tzinis2022optimal}, OCT paired with the refinement of the input-user query to a condition vector which is more amenable to a specified single-conditioned task can fine-tune the conditional model towards the target separation task. Nevertheless, OCT \cite{tzinis2022optimal} lacks the generality of HCT \cite{tzinis22_interspeech} while also performing much worse compared to the optimal condition model. 

 \begin{figure}[t!]
    \centering
    \includegraphics[width=\linewidth]{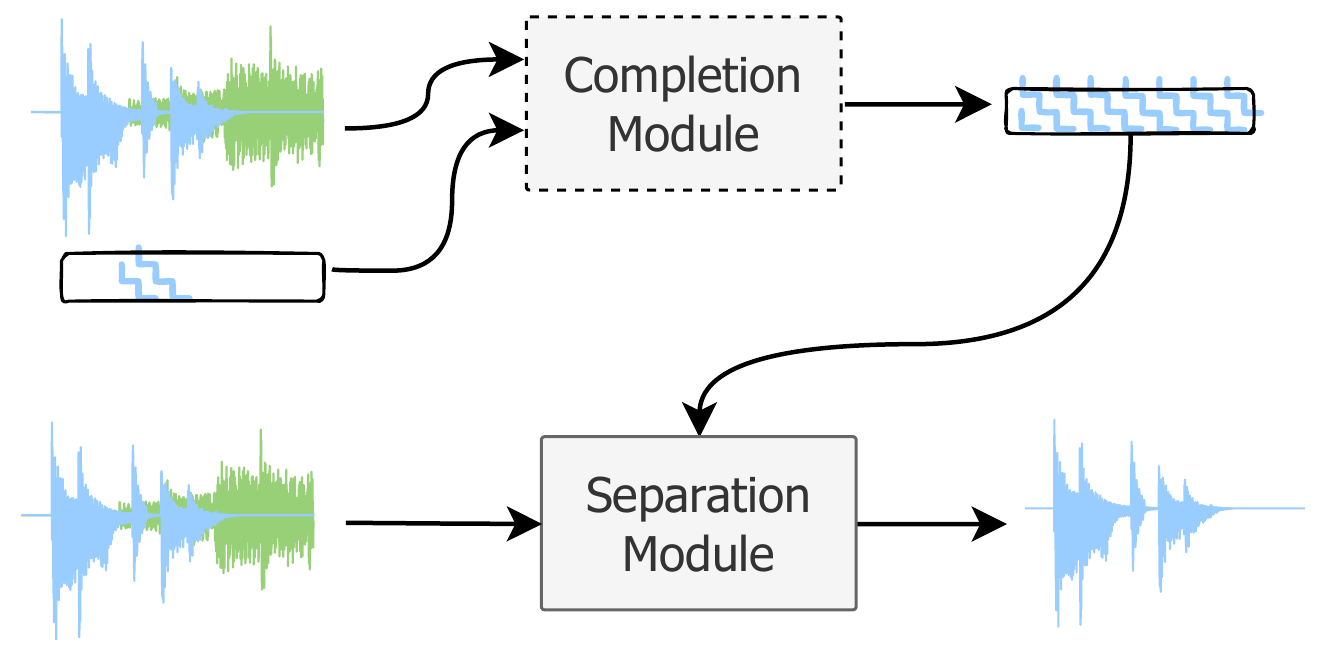}
      \caption{Illustration of our two-stage proposed approach. First, given a mixture and a partial attribute vector describing one of the sources present in the mixture, the completion module is tasked with inferring the rest of the attribute vector. Then, the separation module takes as input the completed condition vector as well as the mixture and is trained to retrieve the target source, while the pre-trained completion module is kept frozen indicated by the dashed outline.}
      \label{fig:twostage}
    \vspace{-15pt}
\end{figure} 

In our work, we focus on how to train multi-conditioned separation systems which are able to perform on-par with dedicated single-conditioned models and how to bridge the performance gap with input-queries that contain the maximal possible description of a target sound source (e.g. for the task of target speech extraction a complete input-query could contain information about the gender, the loudness, the time-order, and the spatial position of the target speaker). Specifically, we explore the scenario where partial utterance-level source attributes are available and we investigate whether estimating missing attributes is beneficial to target speech extraction. We propose a simple yet novel and effective metadata completion mechanism (see Fig.~\ref{fig:twostage}) to estimate the missing parts from an incomplete input-query that describes a target source from the input mixture. At a second step, the estimated completed condition vector is used to train the conditional separation model which is able to access potentially more information associated with the source of interest. Our experiments show that our completion module achieves high accuracy in estimating the missing data for easier and harder datasets. Consequently, our method delivers a much more robust multi-condition separation model scoring on-par with specialist single-conditioned models and close to oracle PIT unconditional methods as well as oracle fully-informed queried models that assume that the user has specified every nuance characteristic of the source of interest.















\section{Method}
\label{sec:method}

\subsection{Task Definition}
\label{ssec:task}

In the general case, the problem of conditional source separation can be formulated in the following way. We consider a mixture waveform $\mathbf{x} = \sum_{i = 1}^N \mathbf{s}_i \in \mathbb{R}^L$ consisting of $N$ sources and a size of $L$ time-samples. We also consider a condition $\mathcal{C}$ (e.g. the amount of energy of a source) and a condition value $v \in \mathcal{C}$ (e.g. low energy or high energy) which describes a target source $\mathbf{s}_T$ where $T = j$ if $\mathcal{C}(\mathbf{s}_j) = v$. The task is then to retrieve the target source $\mathbf{s}_T$ given the mixture $\mathbf{x}$ and the condition value $v$, represented as a vector $\mathbf{c}$.

Additionally, we consider that each source $\mathbf{s}_i$ in the mixture $\mathbf{x}$ is associated with a set of source attributes values. Source attributes can be absolute (i.e., independent of the mixture) or relative (i.e., dependent on the mixture). For instance, the speaker's gender can be thought of as an absolute attribute, while the source energy can be regarded as a relative attribute dependent on the mixture. 

In this work, in order to provide an illustration we narrow our focus on a specific scenario where $N = 2$ and the source attribute set is $\mathscr{C} = \{ \mathcal{G}, \mathcal{E}, \mathcal{O}, \mathcal{S}\}$, where $\mathcal{G}$ is the source attribute that represents the speaker's gender (with values: female/male), $\mathcal{E}$ the energy of the source (high/low), $\mathcal{O}$ the temporal order of appearance of the source (first/second) and $\mathcal{S}$ the spatial distance of the source with respect to the microphone (far/near). Further, we consider conditions $\mathcal{C} \in \mathscr{C}$ meaning each condition $\mathcal{C}$ consists of one source attribute. For example, a valid condition value would be the ``high energy source" or the ``female speaker". Moreover, we assume that every condition value $v$ can be associated only with one source of the mixture, implying that sources have complementary attributes. Thus, each condition value $v$ can be represented as an 8-dimensional one-hot vector $\mathbf{c}$, and the source attribute values of a specific source as an 8-dimensional multi-hot vector $\mathbf{c}_{\text{full}}$. Finally, we assume a fully supervised training scenario where the ground truth sources and their corresponding attributes are available when generating the synthetic training mixtures.

\subsection{Heterogeneous Target Separation}
\label{ssec:hct}

One baseline approach to tackle the problem is to train a conditional separation network using the Heterogeneous Conditional Training (HCT) \cite{tzinis22_interspeech}. In this case, we have a multi-conditional network: 
\begin{equation}
\label{eq:het}
    \begin{gathered}
    \widehat{\mathbf{s}}_T, \, \widehat{\mathbf{s}}_O = f_\theta (\mathbf{x}, \mathbf{c}),
    \end{gathered}
\end{equation}
which is trained to estimate the target $\widehat{\mathbf{s}}_T$ and the residual source $\widehat{\mathbf{s}}_O$ given a one-hot vector $\mathbf{c}$ associated with a condition value $v$. During training, for a given target source $\mathbf{s}_T$ and a uniformly sampled condition value $v$, HCT tries to minimize the following loss:
\begin{equation}
\label{eq:het}
    \begin{gathered}
    \mathcal{L}_{\text{HCT}} = \ell \left(\widehat{\mathbf{s}}_T, {\mathbf{s}}_T \right) + \ell \left(\widehat{\mathbf{s}}_O, {\mathbf{s}}_O \right),
    \end{gathered}
\end{equation}
where $\ell$ is a signal level time-domain loss function. 

\subsection{Proposed Approach}
Our motivation is that some of the target source's attributes can be more discriminative than others, which suggests that we can improve separation performance by ``completing" the user's query.

Consequently, our proposed approach consists of two stages, a completion step followed by a separation step. First, we learn a completion module $g_\phi$, which given the input mixture $\mathbf{x}$ and a sampled condition value $v$ describing the target source $\mathbf{s}_T$, represented as the partial condition vector $\mathbf{c}$, is tasked with retrieving the rest of the source attribute values associated with $\mathbf{s}_T$:  
\begin{equation}
\label{eq:comp}
    \begin{gathered}
    \widehat{\mathbf{c}}_{\text{full}} = g_\phi (\mathbf{x}, \mathbf{c}).
    \end{gathered}
\end{equation}
where $\widehat{\mathbf{c}}_{\text{full}}$ is an 8-dimensional vector consisting of the probabilities assigned by the model to each source attribute value. 

Next, after pre-training $g_\phi$, we aim to learn a conditional separation model $f_\theta$, which given the mixture $\mathbf{x}$ and the concatenation of the condition vector $\mathbf{c}$ and the estimated completed vector $\widehat{\mathbf{c}}_{\text{full}}$, can retrieve the target source $\mathbf{s}_T$ as well as the residual source $\mathbf{s}_O$:
\begin{equation}
\label{eq:sep}
    \begin{gathered}
    \widehat{\mathbf{s}}_T, \, \widehat{\mathbf{s}}_O = f_\theta (\mathbf{x}, \left[\mathbf{c}, \widehat{\mathbf{c}}_{\text{full}} \right]).
    \end{gathered}
\end{equation}
While, the separation model $f_\theta$ is being trained we keep the parameters $\phi$ of completion model $g_\phi$ frozen. By conditioning the separation model both on the original condition vector $\mathbf{c}$ as well as the estimated completion $\widehat{\mathbf{c}}_{\text{full}}$, we enable detection and correction of possibly erroneous estimates from the completion module. 

\section{Experimental Framework}
\label{sec:exp}

\subsection{Dataset}
\label{ssec:datasets}

For our experiments, we construct synthetic mixtures based on single-speaker utterances from Spatial LibriSpeech, as introduced in \cite{tzinis22_interspeech}, which is a synthetically reverberant dataset based on speech utterances from LibriSpeech \cite{panayotov2015librispeech}.

In order to evaluate our method in a range of settings we create an ``Easy" and a ``Hard" dataset as described below. The mixing process consists of first sampling a female and a male speech utterance as well as a near-field and a far-field Room Impulse Response (RIR). By convolving each speech utterance with a randomly matched RIR we get the two source waveforms. Those sources are then mixed with an overlap sampled from $\mathcal{U}[60, 100]\%$ and Signal-to-Noise Ratio (SNR) sampled from $\mathcal{U}([-5, -0.5] \cup [0.5, 5]) \, \text{dB}$ in the case of the Easy dataset and $\mathcal{U}[80, 100]\%$ and $\mathcal{U}([-2.5, -0.5] \cup [0.5, 2.5])\, \text{dB}$ in the case of the Hard dataset. As a result, our generated mixtures consist of two sources $\mathbf{s}_1$ and $\mathbf{s}_2$ each with complementary source attributes (i.e., $\mathcal{G}$: Gender, $\mathcal{E}$ Energy, $\mathcal{O}$: Source order, $\mathcal{S}$: Spatial position) that can be represented in an 8-dimensional multi-one vector. Finally, we sample one condition $\mathcal{C}$ and a condition value $v \in \mathcal{C}$, that describes the target source $\mathbf{s}_T$, which we represent it as an 8-dimensional one-hot vector $\mathbf{c}$.

In total we generate 20,000, 3,000 and 3,000 training, validation and test samples, respectively, sampled at 8 kHz and with a 5-second duration. We note that during each training epoch generation is dynamic and performed on-the-fly.

\subsection{Completion Model}

We construct the completion model based on the ECAPA-TDNN \cite{desplanques20_interspeech} architecture which has shown impressive performance in the domain of speaker verification. In order to be able to inject the conditional information $\mathbf{c}$ to the model, we make a series of modifications to the architecture. First, we add a FiLM \cite{perez2018film} layer right after the residual connection inside every SE-Res2Block. We also add a FiLM layer before the final fully-connected layer of the network, as well as a sigmoid activation after it. 

To reduce the model size of the modified ECAPA-TDNN network, we reduce the number of channels in the convolutional frame layers to 128 (C = 128). The final fully-connected layer consists of four nodes, one for each estimated source attribute, with the output then converted into the 8-dimensional vector $\widehat{\mathbf{c}}_{\text{full}}$, containing the probabilities assigned by the model to each attribute value. The number of model parameters is equal to 0.63M. The network is fed the 64-dimensional log Mel spectrogram of the mixture, which is extracted from its STFT computed with a Hann window of size 256 samples and 50\% overlap. We set the lower cutoff frequency to 50 Hz and we apply 2-dimensional batch normalization \cite{ioffe2015batch} after the log Mel spectrogram extraction as in \cite{kong2020panns}.   

\subsection{Separation Model}

For the conditional separation model we use the conditional Sudo rm -rf  \cite{tzinis2022compute} described in \cite{tzinis22_interspeech, tzinis2022optimal}. Specifically, inside each U-ConvBlock, a FiLM \cite{perez2018film} layer is added right after the residual connection. In an effort to reduce model size, we use $B = 8$ U-ConvBlocks and the number of intermediate channels is set to 512. The number of encoder and decoder trainable bases is set to 512, with each filter having a length of 41 taps and a hop size of 20. The total number of trainable parameters is 5.38M.

\subsection{Training and Evaluation Details}


The completion model is trained using the binary cross entropy loss with a batch size of 6. We use the Adam \cite{adam} optimizer with an initial learning rate equal to $10^{-3}$ which decays to half of its previous value every 40 epochs, along with a weight decay of $2 \cdot 10^{-5}$. Gradient clipping is also performed when the $L_2$-norm is greater than $5.0$. In the experiments involving the ``Easy" (inv. ``Hard") dataset we train the completion model for 50 (inv. 200) epochs. Even though the model has not fully converged at 50 epochs, we show that we can save computation while its performance remains adequate for the downstream task of source separation. 


All of the separation models are trained with the objective to minimize the negative scale-invariant signal to distortion ratio (SI-SDR) \cite{le2019sdr} defined as $\text{SI-SDR}(\widehat{\mathbf{s}}, \mathbf{s}) = 10 \log_{10} \left({\| \alpha \mathbf{s} \|^2}/{\| \alpha \mathbf{s} - \widehat{\mathbf{s}} \|^2} \right)$
where $\alpha$ is a scalar equal to $\alpha = \widehat{\mathbf{s}}^{\top} \mathbf{s} / \| \mathbf{s} \|^2$. A batch size of 6 is used, along with the Adam \cite{adam} optimizer with an initial learning rate equal to $10^{-3}$ decaying to half of its previous value every 20 epochs without any weight decay. We also perform gradient clipping when the $L_2$-norm exceeds $5.0$. Models are evaluated after completing 150 training epochs where we empirically observe model convergence. We evaluate the estimated target source $\widehat{\mathbf{s}}_T$ against the ground truth target source ${\mathbf{s}}_T$ using the SI-SDR.

\section{Results \& Discussion}
\label{sec:results}

\subsection{Missing Attribute Completion}
\label{ssec:completion}

First, we conduct a set of experiments evaluating the performance of the completion models in the Easy and Hard dataset. In Tables~\ref{tab:comp_easy},~\ref{tab:comp_hard} we summarize the results in terms of completion accuracy for every combination of condition $\mathcal{C} \in \mathscr{C}$ and source attributes. The completion model given the mixture $\mathbf{x}$ and a condition vector $\mathbf{c}$, in the form of a one-hot vector, is tasked with estimating the rest of the source attribute values corresponding to the same target source.

Overall, we note that the model achieves impressive completion accuracy in both datasets. We observe that some attributes are easier to predict that others, for instance gender and spatial position, which are absolute attributes, are easier to predict compared to source energy and source order, which are relative instead. A possible explanation could be that in order to predict relative attributes the completion module is required to perform some degree of semantic separation and identification of the sources, rendering the prediction task more difficult. Furthermore, we note that there is a rough symmetry in the accuracy matrix; that is, the model can predict $\mathcal{O}$ given $\mathcal{S}$ roughly as well as predict $\mathcal{S}$ given $\mathcal{O}$.

Comparing the performance between the two datasets, we first notice that even though we train for fewer epochs on the Easy dataset the model accuracy is very high. In the case of the Hard dataset, we mostly observe an expected drop in performance in the relative attributes of source energy $\mathcal{E}$ and source order $\mathcal{O}$. However, the accuracy of $\mathcal{G}$ and $\mathcal{S}$ pairs remains unaffected.

\begin{table}[!tb]
    \caption{Prediction accuracy of the completion module in the Easy dataset (at least 60\% overlap and 5 dB maximum SNR). Each row corresponds to a different input to the completion model, which is tasked with predicting the source attributes $\mathcal{G}$, $\mathcal{E}$, $\mathcal{O}$ and $\mathcal{S}$. }
    \label{tab:comp_easy}
    \centering
    \sisetup{
    detect-weight, 
    mode=text, 
    tight-spacing=true,
    round-mode=places,
    round-precision=1,
    table-format=2.1
    }
    \begin{tabular}{lllll}
\toprule
& \multicolumn{4}{c}{Completion Accuracy (Easy)} \\   \cmidrule(lr){2-5}
&  $\mathcal{G}$ & $\mathcal{E}$ & $\mathcal{O}$ & $\mathcal{S}$ \\ \midrule
Given $\mathcal{G}$ & - & 87.5\% & 92.6\% & 97.5\% \\ 
Given $\mathcal{E}$ & 88.1\% & - & 83.8\% & 89.2\% \\ 
Given $\mathcal{O}$ & 93.0\% & 85.0\% & - & 94.6\% \\ 
Given $\mathcal{S}$ & 98.1\% & 88.9\% & 93.6\% & - \\ \bottomrule
\end{tabular}
\vspace{-5pt}
\end{table}

\begin{table}[!tb]
    \caption{Prediction accuracy of the completion module in the Hard dataset (at least 80\% overlap and 2.5 dB maximum SNR). }
    \label{tab:comp_hard}
    \centering
    \sisetup{
    detect-weight, 
    mode=text, 
    tight-spacing=true,
    round-mode=places,
    round-precision=1,
    table-format=2.1
    }
     \begin{tabular}{lllll}
\toprule
& \multicolumn{4}{c}{Completion Accuracy (Hard)} \\   \cmidrule(lr){2-5}
& $\mathcal{G}$ & $\mathcal{E}$ & $\mathcal{O}$ & $\mathcal{S}$  \\ \midrule
Given $\mathcal{G}$ & - & 76.8\% & 90.8\% & 98.7\% \\ 
Given $\mathcal{E}$ & 80.8\% & - & 73.6\% & 76.8\% \\ 
Given $\mathcal{O}$ & 91.2\% & 74.1\% & - & 93.3\% \\ 
Given $\mathcal{S}$ & 99.1\% & 73.1\% & 93.6\% & - \\ \bottomrule
\end{tabular}
    \vspace{-15pt}
\end{table}

\subsection{Conditional Separation}
\label{ssec:separation}

\begin{table*}[t]
    \caption{Mean test SI-SDR (dB) separation performance on both the ``Easy" and the ``Hard" partitions of the spatial LibriSpeech dataset for conditional models which are evaluated using different input-condition vectors. $^*$ denotes an oracle ensemble which contains specialist single-conditioned models trained and evaluated on the corresponding condition vector.}
    \label{tab:sep_easy}
    \centering
    \sisetup{
    detect-weight, 
    mode=text, 
    tight-spacing=true,
    round-mode=places,
    round-precision=1,
    table-format=2.1
    }
    {
    \begin{tabular}{lccccc|ccccc}
    \toprule
    & \multicolumn{5}{c}{``Easy" partition} & \multicolumn{5}{c}{``Hard" partition} \\   \cmidrule(lr){2-6} \cmidrule(lr){7-11}
    \multicolumn{1}{l}{Input condition vector} & $\mathcal{G}$ & $\mathcal{E}$ & $\mathcal{O}$ & $\mathcal{S}$ & Overall & $\mathcal{G}$ & $\mathcal{E}$ & $\mathcal{O}$ & $\mathcal{S}$ & Overall  \\ \midrule
    HCT \cite{tzinis22_interspeech} & 11.9 & 11.2 & 8.9 & 11.8 & 10.9 & 10.9 & 9.8 & 6.8 & 10.9 & 9.6 \\
    Compeltion (Ours) & 12.5 & 11.7 & 11.1 & 12.7 & 12.0 & 11.5 & 9.8 & 9.8 & 11.6 & 10.7 \\ \midrule
    Single-condition ensemble$^*$ & 13.0 & 10.2 & 8.5 & 12.6 & 11.1 & 11.8 & 4.9 & 2.1 & 11.6 & 7.6 \\ \midrule
    PIT Oracle & & & & & 12.7 & & & & & 12.0 \\
    Completion Oracle & & & & & 13.1 & & & & & 12.1 \\
    \bottomrule
    \end{tabular}
    }
    \vspace{-5pt}
\end{table*}

In this core set of experiments, we assess our proposed approach in the task of conditional separation as formulated in Section~\ref{ssec:task}. In brief, the module given the mixture $\mathbf{x}$ and a one-hot condition vector $\mathbf{c}$, is tasked with recovering the corresponding target source $\mathbf{s}_T$. We compare against a number of baselines, with the main one being HCT as described in Section~\ref{ssec:hct}, which is a multi-condition model. A second baseline consists of all the single-condition models which are trained and evaluated conditioned only on one source attribute. Subsequently, we evaluate in relation to two oracle models. The first entails Permutation Invariant Training (PIT) \cite{Yu2017PIT} followed by oracle selection of the output source that best matches the target source. The second is the completion oracle trained with the complete attribute vector, which is equivalent to training the conditional separation model with a perfect completion module. 

In Table~\ref{tab:sep_easy} we summarize the results in terms of separation performance in the Easy and Hard datasets, respectively. First, we note that the oracle models outperform almost all of the models, with the completion oracle providing a boost over the PIT oracle, suggesting that the presence of conditional information enhances performance. As we shift our focus to single-condition models, we observe that certain attributes when employed as conditions exhibit consistently better performance, indicating that they are more discriminative compared to the others. For instance, conditioning on gender $\mathcal{G}$ yields remarkable results, comparable to those of the completion oracle and even surpassing those of the PIT oracle in the Easy setting. Nevertheless, other single-condition models employing relative conditions, such as source order $\mathcal{O}$ and source energy $\mathcal{E}$, face challenges in terms of performance, especially in the Hard setting. While, the multi-condition model HCT appears to be able to address this performance degradation to some extent, there is still potential for further improvement, as suggested by the top performing models. This indicates that the incorporation of a completion approach may be necessary or beneficial to enhance the performance of separation.

Shifting our attention to the perfomance of our proposed approach, we observe that it outperforms the HCT multi-condition model in every scenario, with an overall margin of $1.1$ dB in both settings, significantly reducing the gap to the oracle models. Our method demonstrates an improvement in performance for every type of given attribute, especially low performing ones, such as source order $\mathcal{O}$. Moreover, it drastically outperforms the single-condition models based on relative attributes, while matching or approaching the performance of the specialized signle-conditioned models based on absolute attributes. Finally, in Fig.~\ref{fig:scatterplot} we compare the performance of our approach to the HCT model by plotting the performance of each test set sample. The results presented in the figure illustrate the improvement in the mean performance achieved by our approach as compared to the HCT model. Additionally, it shows a reduction in the number of samples with negative SI-SDR, which implies that our approach has the potential to reduce the number of erroneous target source selections.

 \begin{figure}[htb]
    \centering
    \includegraphics[width=0.85 \linewidth]{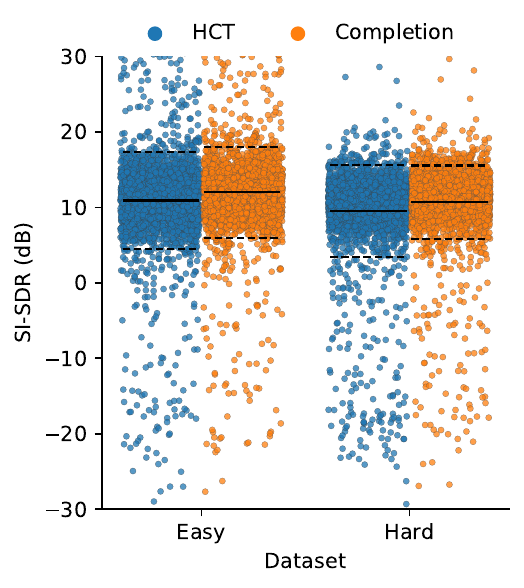}
      \caption{Comparison of the performance of our completion approach with the HCT model on the test set samples. The mean and standard deviation are represented by solid black lines and dashed lines, respectively. Notice that our approach achieves an improvement both in terms of mean performance and in terms of a significant reduction in the number of samples with really low SI-SDR, indicating a potential reduction in erroneous target source selections.}
      \label{fig:scatterplot}
    \vspace{-15pt}
\end{figure}

\section{Conclusion}
\label{sec:conclusion}

In conclusion, we have presented a two-stage approach for conditional source separation with heterogeneous attributes. The first stage involves a completion module responsible for retrieving the missing attributes associated with the target source, given a partial condition vector and the input mixture. The second stage involves a separation module that takes as input the concatenation of the completed and original condition vectors as well as the input mixture and is trained to retrieve the target source. This proposed approach outperforms the previous state-of-the-art multi-condition separation method of heterogeneous condition training by leveraging the discriminative power of the different attributes associated with the target source. Our method could potentially enable users to complete their queries and obtain more accurate and informative separation results.

\bibliographystyle{IEEEtran}
\bibliography{main}

\end{sloppy}
\end{document}